\documentstyle[11pt,newpasp,twoside,epsf]{article}
\def\edcomment#1{\iffalse\marginpar{\raggedright\sl#1\/}\else\relax\fi}
\reversemarginpar

\begin {document}
\title {Spin Rates and Magnetic Fields of Millisecond Pulsars}
 \author {Frederick K. Lamb and Wenfei Yu}
\affil {Center for Theoretical Astrophysics, Department of
  Physics, and Department of Astronomy, University of Illinois at
  Urbana-Champaign, 1110 W. Green Street, Urbana, IL 61801-3080,
  USA}

\begin {abstract}
Observations made using the \textit {Rossi X--ray Timing Explorer} have shown that accreting weak-field neutron stars in low-mass X-ray binary systems (LMXBs) produce three distinct types of millisecond X-ray oscillations that can be used to determine the spin rates and estimate the magnetic fields of these stars. These oscillations show that more than two dozen neutron stars in LMXBs have spin rates and magnetic fields in the range that will make them radio-emitting millisecond pulsars when accretion ceases, supporting the hypothesis that neutron stars in LMXBs are the progenitors of the rotation-powered millisecond pulsars. The spins of the 16 known accreting millisecond pulsars in LMXBs are consistent with spin-up to accretion spin equilibrium for magnetic fields ranging from $3 \times 10^7$~G to $3 \times 10^8$~G and time-averaged accretion rates ranging from $3 \times 10^{-3}{\dot M}_E$ to ${\dot M}_E$, provided these stars have been accreting long enough. The $P$--${\dot P}$ distribution of rotation-powered millisecond pulsars indicates that their initial spins are set by spin-up to spin equilibrium at accretion rates ranging from $10^{-3}{\dot M}_E$ to ${\dot M}_E$ or that they never reached accretion spin equilibrium.
\end {abstract}

\section {Introduction}

Soon after the discovery of millisecond pulsars (Backer et al.\ 1982), it was suggested (Alpar et al.\ 1982; Radhakrishnan and Srinivasan 1982) that they have been recycled by being spun up by accretion of angular momentum in low-mass X-ray binary systems (LMXBs). Although this picture has been widely accepted for more than two decades (see Bhattacharya 1995), until the last few years there was only indirect evidence that neutron stars in LMXBs have the spin rates and magnetic fields required for them to be the progenitors of the rotation-powered millisecond pulsars (MSPs). The \textit {Rossi X-ray Timing Explorer} (\textit {RXTE}) has transformed the situation by revealing that accreting neutron stars in LMXBs produce three distinct types of millisecond X-ray oscillations that can be used to determine their spins and estimate their magnetic fields. All three types of oscillation are thought to be generated directly or indirectly by the star's magnetic field and spin. Together, they provide strong evidence that more than two dozen neutron stars in LMXBs have the millisecond spin periods and $\sim\,$10$^{8}$--10$^{10}$~G magnetic fields necessary for them to become rotation-powered MSPs when accretion ceases.

\begin {table}[t!]
\begin {center}
\begin {minipage}{122 mm}
\caption {Accretion- and Nuclear-Powered Millisecond Pulsars}
\vspace {3pt}
\begin {tabular}{cll}
\tableline
\tableline
\noalign{\kern 4pt}
$\nu_{\rm spin}$~(Hz)$^a$   &Object   &Reference \\
\noalign{\kern 4pt}
\tableline
\noalign{\kern 3pt}
619\ \ N,K\quad\qquad                 &\hbox{4U~1608$-$52}    & Hartman et al.\ 2003\\
601\ \ N,K\quad\qquad                 &\hbox{SAX~J1750.8$-$2900}    & Kaaret et al.\ 2002\\
589\ \ N\qquad\qquad            &\hbox{X~1743$-$29}     & Strohmayer et al.\ 1997\\
581\ \ N,K\quad\qquad         &\hbox{4U~1636$-$53}    & Zhang et al.\ 1996;\\
                          &                       & Strohmayer et al.\ 1998\\
567\ \ N\qquad\qquad      &\hbox{X~1658$-$298}    & Wijnands et al.\ 2001\\
549\ \ N,K\quad\qquad                 &\hbox{Aql~X-1}         & Zhang et al.\ 1998\\
524\ \ N,K\quad\qquad                 &\hbox{KS~1731$-$260}   & Smith et al.\ 1997\\
\noalign{\kern 4pt}
\tableline
\noalign{\kern 4pt}
435\ \ A\qquad\qquad     &\hbox{XTE~J1751$-$305}    &Markwardt et al.\ 2002\\
\noalign{\kern 1pt}
410\ \ N\qquad\qquad          &\hbox{SAX~J1748.9$-$2021}    &Kaaret et al.\ 2003\\
\ 401\ \ A,N,K\           &\hbox{SAX J1808.4$-$3658\quad}    &Wijnands \& van der Klis 1998;\\
                              &                             &Chakrabarty \& Morgan 1998\\
\noalign{\kern 1pt}
363\ \ N,K\quad\qquad                     &\hbox{4U~1728$-$34}       &Strohmayer et al.\ 1996\\
330\ \ N,K\quad\qquad     &\hbox{4U~1702$-$429}      &Markwardt et al.\ 1999\\
314\ \ A,N\quad\qquad     &\hbox{XTE~J1814$-$338}    &Markwardt et al.\ 2003b\\
270\ \ N\qquad\qquad                         &\hbox{4U~1916$-$05}       &Galloway et al.\ 2001\\
191\ \ A,K\quad\qquad        &\hbox{XTE~J1807.4$-$294}    &Markwardt et al.\ 2003a, 2004\\
185\ \ A\qquad\qquad     &\hbox{XTE~J0929$-$314}\ \ \ \    &Galloway et al.\ 2002\\
\noalign{\kern 4pt}
\tableline
\noalign{\kern 4pt}
\end {tabular}
\end {minipage}
\begin {minipage}{120 mm}
{$^a$Spin frequency inferred from periodic or nearly periodic X-ray oscillations. A: accretion-powered millisecond pulsar. N: nuclear-powered millisecond pulsar. K: kilohertz QPO source. See text for details.}
\end {minipage}
\end {center}
\end {table}

Periodic accretion-powered X-ray oscillations with millisecond periods have been detected in five neutron stars in LMXBs (see Table~1), establishing beyond any doubt that these stars have dynamically important magnetic fields. Their coherent oscillations indicate that they have field strengths $\ga10^{7}$~G (see Miller, Lamb, \& Psaltis 1998, hereafter MLP98), while their nearly sinusoidal waveforms and unusually low oscillation amplitudes indicate that they have field strengths $\la10^{10}$~G (Psaltis \& Chakrabarty 1999). The spin frequencies of these accretion-powered MSPs range from 185~Hz to 435~Hz.

Nearly periodic X-ray oscillations have been detected during thermonuclear bursts of 13 neutron stars in LMXBs, including 2 of the 5 known accretion-powered MSPs (Table~1). The existence of thermonuclear bursts indicates field strengths $\la10^{10}$~G (Lewin, van Paradijs, \& Taam 1995) while the spectra of the persistent X-ray emission (Psaltis \& Lamb 1998) and the properties of the burst oscillations (see \S\,2) indicate field strengths $\ga10^{7}$~G. The spin frequencies of these nuclear-powered MSPs range from 270~Hz to 619~Hz. 

Pairs of accretion-powered kilohertz quasi-periodic X-ray oscillations have been detected in more than two dozen accreting neutron stars (see Lamb 2003), including 8 nuclear-powered MSPs and 2 of the 5 known accretion-powered MSPs (Table~1). The frequencies of these kilohertz QPOs range from $\sim\,$100~Hz to $\sim\,$1300~Hz, showing that accreting gas orbits close to the surfaces of these stars and that they have magnetic fields $\sim 10^{7}$--$10^{9}$~G (MLP98). The separation of the two QPOs remains constant to within a few tens of Hz as their frequencies vary by as much as a factor of 5 and is commensurate with the spin frequency of the star (see \S\,2), showing that the star's spin plays a central role in generating the QPO pair and indicating that the magnetic fields of these stars are $\ga10^{8}$~G. The spin frequencies of the stars that produce kilohertz QPOs can be inferred directly from the frequency of their periodic X-ray oscillations, if they have been detected, or indirectly from the separation of their kilohertz QPOs, if they have not. The spins of the neutron stars in which periodic X-ray oscillations and kilohertz QPOs have both been detected range from 191~Hz to 619~Hz (Table~1).

These discoveries have established that many neutron stars in LMXBs have magnetic fields and spin rates similar to those of the rotation-powered MSPs. The similarity of these stars to rotation-powered MSPs strongly supports the hypothesis that they are the progenitors of rotation-powered MSPs. After being spun down by rotation-powered emission, the neutron stars in these systems are spun up to millisecond periods by accretion of matter from their binary companions, eventually becoming nuclear- and accretion-powered MSPs and then, when accretion ends, rotation-powered MSPs.

In $\S\,$2 we describe in more detail the new evidence that many neutron stars in LMXBs have millisecond spin periods and dynamically important magnetic fields. In $\S\,$3 we discuss the spin evolution of weak-field neutron stars in LMXBs, including the long-standing idea that their spin rates are limited by coupling of their magnetic fields to the surrounding accretion disk and recent proposals that their spin rates may be affected by gravitational radiation by the spinning star. In $\S\,$4 we summarize the implications of the new evidence for the evolution of neutron stars in LMXBs and formation of rotation-powered MSPs via recycling.

\section {Measuring the Spin Rates of Accreting Weak-Field Neutron Stars}

Neutron stars in LMXBs are accreting gas from a Keplerian disk fed by a low-mass companion star. The star's magnetic field and accretion rate are thought to be the most important factors that determine the accretion flow pattern near it and the spectral and temporal characteristics of its X-ray emission (see MLP98). The accretion rates of these stars vary with time and can range from the Eddington critical rate ${\dot M}_E$ to less than $10^{-4}{\dot M}_E$. Their magnetic fields are thought to range from $10^{11}$~G down to $10^7$~G or possibly less. Magnetic fields at the upper end of this range are strong enough to terminate the Keplerian disk well above the stellar surface, even for accretion rates $\sim {\dot M}_E$, whereas magnetic fields at the lower end of this range affect the flow only close to the star, even for accretion rates as low as $\sim 10^{-4}{\dot M}_E$.

For intermediate field strengths and accretion rates, some of the accreting gas is expected to couple to the star's magnetic field well above the stellar surface and be funneled toward the magnetic poles, heating the stellar surface unevenly. The remainder of the accreting gas is expected to remain in a geometrically thin Keplerian flow that penetrates close to the stellar surface, as shown in Figure~1. This flow is thought to be responsible for generating the kilohertz QPOs (see Lamb \& Miller 2001; Lamb 2003; Lamb \& Miller 2004). Recent observations (see below) show that X-ray bursts also heat the stellar surface unevenly, and that the regions heated in this way are coupled to the rotation of the star, probably via the star's magnetic field. Whether due to accretion or to nuclear burning, uneven heating of the stellar surface produces a broad pattern of X-ray emission. Rotation of this pattern makes both the accretion-powered and nuclear-powered X-ray emission of the star appear to oscillate at its spin frequency.

\begin {figure}[t]
  \plotfiddle{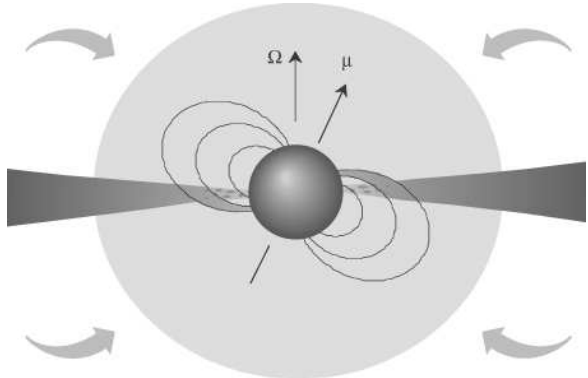}{1.8 in}{0}{50}{50}{-110}{0}
  \caption {Side view of a weak-field neutron star accreting from a disk, showing the complex flow pattern expected. Some accreting gas couples strongly to the magnetic field and is funneled toward the magnetic poles, but a substantial fraction couples only weakly and drifts inward in nearly circular orbits as it transfers its angular momentum to the star via the stellar magnetic field. From~MLP98.}
  \end {figure}

\begin {figure}[t]
  \plotfiddle{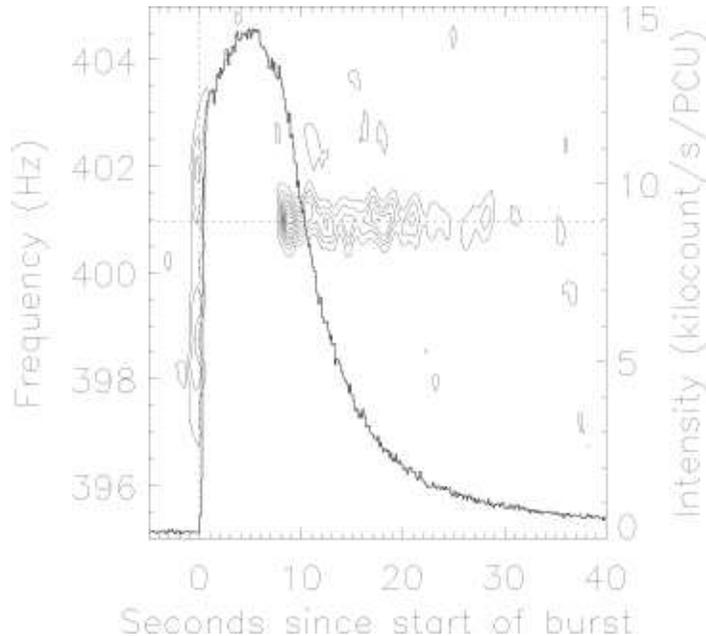}{3.2 in}{0}{75}{75}{-130}{-0}
  \caption {An X-ray burst from \mbox {SAX~J1808.4$-$3658} observed on 18 October 2002. \textit {Dark curve and scale at right}: X-ray count rate as a function of time during the burst. \textit {Contours and scale at left}: Dynamic power spectrum of the X-ray brightness, showing the rapid increase in the frequency of the burst oscillation from 397~Hz to 403~Hz during the rise of the burst, the disappearance of the oscillation at the peak of the burst, and its reappearance about 10~s after the start of the burst. The horizontal dashed line shows the frequency of the neutron star's spin inferred from its accretion-powered periodic X-ray oscillations. From Chakrabarty et al.~(2003).}
\end {figure}

A breakthrough was achieved during the past year with the detection of burst oscillations in the accretion-powered MSPs \mbox {SAX~J1808.4$-$3658} (Chak\-rabarty et al.\ 2003) and \mbox {XTE~J1814$-$338} (Strohmayer et al.\ 2003) and precise measurements of the frequencies and phases of these oscillations. The new results show that, except during the first seconds of some bursts, \textit {the burst oscillations of these stars have the same frequency and phase as their coherent accretion-powered oscillations} (see Fig.~2), establishing beyond any doubt that (1)~these stars have magnetic fields strong enough to channel the accretion flow and enforce corotation of the gas heated by nuclear burning and (2)~the nuclear- and accretion-powered oscillations are both produced by spin modulation of the X-ray flux from the stellar surface. The burst oscillations of some other stars are very stable (Strohmayer \& Markwardt 2002), but many show frequency drifts and phase jitter (Strohmayer et al.\ 1996; Strohmayer et al.\ 1998; Muno, Fox, \& Morgan 2000; Muno et al.\ 2002). The new results confirm that burst and persistent oscillations both reveal directly the spin frequency of the star.

The 16 MSP spins measured to date are consistent with a uniform distribution that ends abruptly at 760~Hz (Chakrabarty et al.\ 2003), but they are also consistent with a distribution that decreases more gradually. The proportion of accretion- and nuclear-powered MSPs with frequencies higher than 500~Hz is greater than the proportion of known rotation-powered MSPs with such high frequencies, probably because there is no bias against detecting accretion-powered MSPs with high frequencies, whereas detection of rotation-powered MSPs with high frequencies is still difficult (see Chakrabarty et al.\ 2003). 

Seven of the 16 known MSPs in LMXBs have frequencies $>435$~Hz, whereas none of the accretion-powered MSPs have frequencies this high (see Table~1). Although the current sample is too small to draw a conclusion, \textit {such a trend is to be expected if many of these MSPs are near accretion spin equilibrium, because pulsars with stronger magnetic fields should produce stronger oscillations and have lower equilibrium spin rates, other things being equal (see \S\,3)}.

Classic kilohertz QPO pairs have been discovered in two accretion-powered MSPs, XTE~J1807.4$-$294 (Markwardt et al.\ 2003a) and SAX~J1808.4$-$3658 (Wijnands et al.\ 2003; see Fig.~3). The frequency separation of the QPO pair is consistent with the spin frequency of XTE~J1807.4 but with half the spin frequency of SAX~J1808.4. The kilohertz QPO separation is consistent with the spin frequency or half of it in all stars in which burst oscillations have been detected (see Lamb \& Miller 2004; Lamb 2005). These discoveries demonstrate conclusively that some kilohertz QPO sources have dynamically important magnetic fields and that the spin of the star plays a central role in generating the QPO pair, confirming two important predictions of the sonic-point beat-frequency model (MLP98). But they also show that the original model is incorrect or at least incomplete, because it cannot explain a frequency separation equal to \textit {half} the spin frequency (Lamb \& Miller 2004). 

A modified version of the sonic-point beat-frequency model (Lamb \& Miller 2004; Lamb 2005) attributes the lower kilohertz QPO to interaction of X-rays from the stellar surface with vertical motions of gas in the disk excited by the star's radiation or magnetic field at the radius where its spin frequency resonates with the vertical epicyclic frequency. The lowest-order linear resonance of this type would generate vertical oscillations with half the spin frequency and could produce QPO frequency separations equal to the star's spin frequency or half of it (Lamb \& Miller 2004; Lamb 2005).

The new results exclude models (see Lamb 2003) in which the frequencies of the kilohertz QPOs are various relativistic precession frequencies because spin frequencies several times higher than those observed would be required and because such models cannot explain the commensurability of the frequency separation of the QPOs with the stellar spin frequency (Lamb 2003; Lamb 2005).

\begin {figure}[t]
  \plotfiddle{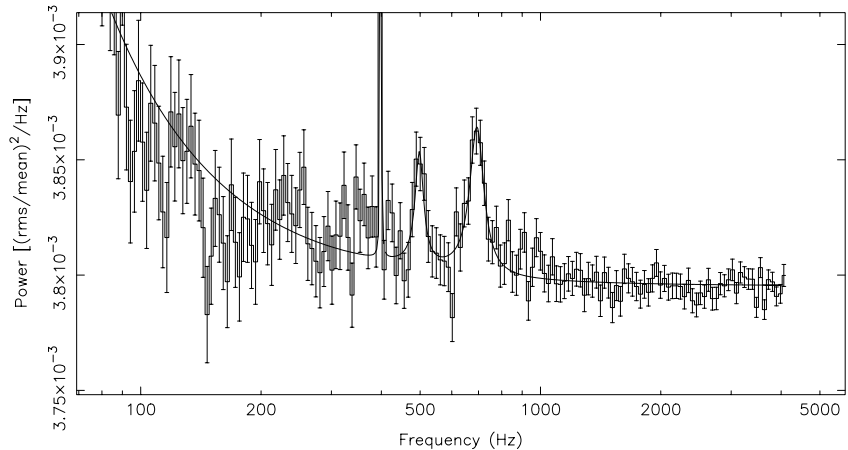}{2.1 in}{0}{45}{45}{-170}{+45}
 \caption {Power density spectrum of the X-ray brightness variations of the accretion-powered MSP \mbox {SAX~J1808.4$-$3658} on 18 October 2002, showing the 401~Hz periodic oscillations (``pulsations'') at the star's spin frequency, the lower kilohertz QPO at 499$\pm4$~Hz and the upper kilohertz QPO at 694$\pm4$~Hz (Wijnands et al.\ 2003).}
\end {figure}

\section {Evolution of Neutron Star Spins and Production of Millisecond Rotation-Powered Pulsars in LMXBs}

\textit {Current spins of neutron stars in LMXBs}.---Accretion from a disk will spin up a slowly-rotating neutron star on the spin-relaxation timescale (Ghosh \& Lamb 1979, hereafter GL79; Ghosh \& Lamb 1992, hereafter GL92; Lamb \& Yu 2004)
\begin {equation}
t_{\rm spin} \equiv 2\pi\nu_{\rm spin}I/[{\dot M} (GMr_m)^{1/2}] \sim 10^8 \, {\rm yr}\, \left( \frac{\nu_{\rm spin}}{\rm 300~Hz} \right)  \left( \frac{\dot M} {0.01{\dot M}_E} \right)^{-1+\alpha/3}\;,
\end {equation}
where $\nu_{\rm spin}$, $M$, and $I$ are the star's spin, mass, and moment of inertia, ${\dot M}$ is the accretion rate onto the star (not the mass transfer rate), $r_m$ is the angular momentum coupling radius, $\alpha$ is 0.23 if the inner disk is radiation-pressure-dominated (RPD) or 0.38 if it is gas-pressure-dominated (GPD), and in the last expression on the right the weak dependence of $t_{\rm spin}$ on $M$, $I$, and the star's magnetic field has been neglected.

The current spin rates of neutron stars in LMXBs reflect the average accretion torque over a period $\sim t_{\rm spin}$ or longer. Determining this average torque is complicated by the fact that the accretion rates and magnetic fields of these stars vary with time by large factors and that the torque can decrease as well as increase the spin rate. While a few neutron stars in LMXBs accrete steadily at rates $\sim {\dot M}_E$, most accrete at rates $\sim 10^{-3}$--$10^{-2}{\dot M}_E$ (Hasinger \& van der Klis 1989; Lamb 1989; van den Heuvel 1992; MLP98) and many accrete only episodically (van den Heuvel 1992; Ritter \& King 2001). The recently-discovered accretion-powered MSPs in LMXBs have outbursts only every few years, during which their accretion rates rise to $\sim 10^{-2}{\dot M}_E$ for a few weeks before falling again to $\la 10^{-4}{\dot M}_E$ (see Chakrabarty et al.\ 2003; Strohmayer et al.\ 2003). Also, there is strong evidence that the magnetic fields of neutron stars in LMXBs decrease by factors $\sim 10^2$--$10^3$ during accretion, perhaps on timescales as short as hundreds of years (see Shibazaki et al.\ 1999; Bhattacharya \& Srinivasan 1995).

If a star's magnetic field and accretion rate are constant, accretion will spin it up on a timescale $\sim t_{\rm spin}$ to its equilibrium spin frequency $\nu_{\rm eq}$. This frequency depends on $M$, the strength and structure of the star's magnetic field, the thermal structure of the disk at $r_m$, and ${\dot M}$ (GL79; White \& Stella 1987; GL92). If a star's magnetic field and accretion rate change on timescales longer than $t_{\rm spin}$, the spin frequency will approach $\nu_{\rm eq}$ and track it as it changes. If instead ${\dot M}$ varies on timescales shorter than $t_{\rm spin}$, the spin rate will fluctuate about the appropriate average value of $\nu_{\rm eq}$ (see Elsner, Ghosh, \& Lamb 1980). Thus $\nu_{\rm eq}$ and its dependence on $B$ and ${\dot M}$ provide a framework for analyzing the evolution of the spins of neutron stars in LMXBs. 

Figure~4 shows $\nu_{\rm eq}$ for five accretion rates and dipole magnetic fields $B_d = 3.2 \times 10^{19} (P{\dot P})^{1/2}$~G ranging from $10^{7}$~G to $10^{11}$~G. The lines are actually bands, due to systematic uncertainties in the models. The lines for ${\dot M}={\dot M}_E$ and ${\dot M}=0.1{\dot M}_E$ have jumps where the structure of the disk at the angular momentum coupling radius $r_m$ changes from RPD (lower left) to GPD (upper right); in reality the transition is smooth. For ${\dot M} \la 0.01{\dot M}_E$, the disk is GPD at $r_m$ even if the star's magnetic field is $\la 3 \times 10^7$~G. The effects of the stellar surface and the innermost stable circular orbit (Lamb \& Yu 2004) are not shown. 

The properties of the 16 known MSPs in LMXBs (Table~1) are consistent with spin-up by accretion to spin equilibrium if their magnetic fields are between $3 \times 10^7$~G and $3 \times 10^8$~G and their time-averaged accretion rates are between $3 \times 10^{-3}{\dot M}_E$ and ${\dot M}_E$. The spin rates and visible pulsations of the accretion-powered MSPs are understandable if they have magnetic fields $\sim 3 \times 10^8$~G and have been spun up to accretion spin equilibrium by accretion at rates \mbox {$\sim 10^{-2} {\dot M}_E$}. The higher spin rates of the other MSPs are understandable if they have magnetic fields $\la 10^8$~G and time-averaged accretion rates $\sim 10^{-2}{\dot M}_E$. The absence of MSPs with spin frequencies $>750$~Hz is consistent with spin-up to accretion spin equilibrium if these stars have magnetic fields $\ga 3 \times 10^7$~G and average accretion rates \mbox{$\la 10^{-3}{\dot M}_E$}. These fields and rates are consistent with the other observed properties of individual neutron stars in LMXBs (MLP98; Psaltis \& Chakrabarty 1999; Chakrabarty et al.\ 2003). Alternatively, these stars may have stopped accreting before reaching spin equilibrium or been spun down as accretion ended.

\begin {figure}[t!]
\plotfiddle{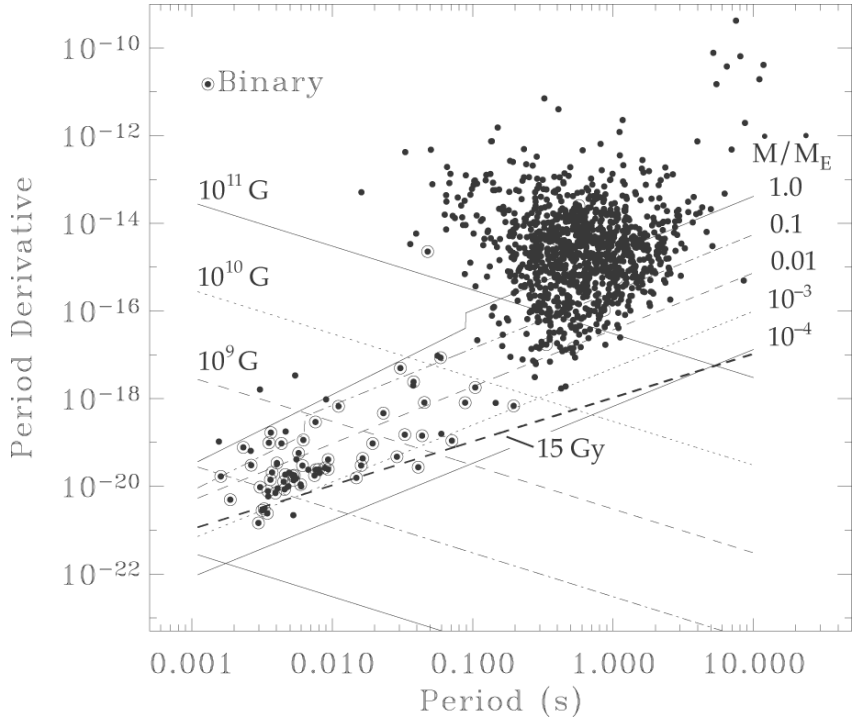}{3.5 in}{0}{72}{72}{-166}{-10}
  \caption {Spin-evolution diagram. Lines sloping downward to the right show the $P$-$\dot P$ relation for magnetic dipole braking by a field with the strength indicated. Lines sloping upward to the right show the equilibrium spin period of a neutron star with the accretion rate indicated by the labels and a dipole field of the strength indicated by the downward-sloping lines. The dashed line  sloping upward to the right shows where stars with a spin-down time equal to 15~Gy would lie. Data points are known rotation-powered pulsars; those of pulsars in binary systems are encircled. Data from Hobbs \& Manchester (2004).}
\end {figure}

Based on the limited information then available, some authors (Bildsten 1998; Ushomirsky, Cutler, \& Bildsten 2000) speculated that neutron stars in LMXBs have negligible magnetic fields and spin frequencies in a narrow range, with many within 20\% of 300~Hz. Such a distribution would be difficult to explain by accretion torques and was taken as evidence that gravitational radiation plays an important role. We now know (see \S\,2) that most if not all neutron stars in LMXBs have dynamically important magnetic fields, that the observed spins of neutron stars in LMXBs are approximately uniformly distributed from $<200$~Hz to $>600$~Hz, and that production of gravitational radiation by uneven heating of the crust or excitation of $r$-waves is not as easy as was originally thought (Ushomirsky et al.\ 2000; Lindblom \& Owen 2002). The spins of neutron stars in LMXBs may be affected by gravitational radiation, if their magnetic fields are weak enough, but there is at present no evidence for this.

\textit {Production of millisecond rotation-powered pulsars}.---The initial proposals that neutron stars are spun up to millisecond periods by accretion in LMXBs assumed that they accrete at rates $\sim {\dot M_E}$ throughout their accretion phase (see Bhattacharya \& van den Heuvel 1991) and implicitly that accretion then ends suddenly; otherwise the stars would track $\nu_{\rm eq}$ to low spin rates as accretion ends. This simplified picture is sometimes still used (see, e.g., Arzoumanian, Cordes, \& Wasserman 1999), but---as noted above---most neutron stars in LMXBs accrete at rates $\ll {\dot M}_E$ and many accrete only episodically. The real situation is therefore more complex.

The initial spins of rotation-powered MSPs recycled in LMXBs are the spins of their progenitors when they stopped accreting. These spins depend sensitively on the magnetic fields and the appropriately averaged accretion rates of the progenitors when accretion ends. Comparison of the equilibrium spin-period curves for a range of accretion rates with the $P$--${\dot P}$ distribution of known rotation-powered MSPs (Fig.~4) suggests four important conclusions:

(1)~The overall $P$--${\dot P}$ distribution is consistent with the distribution expected for spin-up to accretion spin-equilibrium and generally supports the models of disk accretion by weak-field neutron stars used. In particular, the observed $P$--${\dot P}$ distribution is consistent with the predicted distribution only if the accretion torque vanishes at a spin frequency close to the predicted $\nu_{\rm eq}$.
 
(2)~The accretion spin-equilibrium hypothesis predicts that MSPs should not be found above the spin-equilibrium line for ${\dot M} = {\dot M}_E$ and $\omega_c = 1$, because this is a bounding case. The observed $P$--${\dot P}$ distribution is consistent with the RPD model of the inner disk that was used for ${\dot M}\ga 0.1 {\dot M_E}$, except for two pulsars discovered very recently in globular clusters (\mbox {B1821$-$24} [lower-left] and \mbox {B1820$-$30A} [upper-right]; Hobbs et al.\ 2004). Either the intrinsic $\dot P$'s of these pulsars are lower than shown or the RPD model of the inner disk does not accurately describe the accretion flow that spun up these stars.

(3)~The accretion spin-equilibrium hypothesis predicts that MSPs should be rare or absent below the spin-equilibrium line for ${\dot M} = 10^{-4} {\dot M}_E$, because stars accreting at such low rates generally will not achieve spin equilibrium during their accretion phase. The observed $P$--${\dot P}$ distribution is consistent with this prediction.

(4)~The MSPs near the 15~Gyr spin-down line were produced \textit {in situ} by final accretion rates $\la 3 \times 10^{-3} {\dot M}_E$ rather than by spin-up to shorter periods by accretion at rates $\ga 3 \times 10^{-3} {\dot M}_E$ followed by magnetic braking, because braking would take too long. This result accords with the expectation (see above) that most neutron stars in LMXBs accrete at rates $\ll {\dot M}_E$ toward the end of their accretion phase.

\section {Concluding Remarks}

The \textit {RXTE} mission has discovered 5 accretion-powered and 13 nuclear-powered MSPs, with frequencies ranging from 185~Hz to 619~Hz. Nine of these MSPs produce kilohertz QPOs, and the kilohertz QPOs of another dozen neutron stars in LMXBs indicate that they also have 200--600~Hz spin frequencies. The \textit {RXTE} observations indicate that the MSPs in LMXBs have magnetic fields ranging from $3 \times 10^7$~G to $3 \times 10^8$~G. If they still have magnetic fields this strong and 200--600~Hz spin rates when accretion ends, these neutron stars are likely to become radio-emitting MSPs. These discoveries strongly support the hypothesis that neutron stars in LMXBs are the progenitors of rotation-powered MSPs.

The current spin rates of neutron stars in LMXBs reflect the average of the accretion torque over the current spin-relaxation timescale, which typically is \mbox {$\ga 10^7$~yr}. It is difficult to compute the expected average accretion torque over this period, because the torque-averaged accretion rate is uncertain. If accretion torque theory can be validated, measurements of current spin rates can provide estimates of the average accretion rate over this period to compare with binary evolution calculations. 

The properties of the 16 known accretion- and nuclear-powered MSPs in LMXBs are consistent with spin-up by accretion to spin equilibrium if their magnetic fields are between $3 \times 10^7$~G and $3 \times 10^8$~G and their time-averaged accretion rates are between $3 \times 10^{-3}{\dot M}_E$ and ${\dot M}_E$. The spins of neutron stars in LMXBs may be affected by gravitational radiation if their magnetic fields are weak enough, but there is at present no evidence for this. 

The initial spins of recycled rotation-powered MSPs reflect their magnetic fields and accretion rates when accretion ended. The $P$--${\dot P}$ distribution of rotation-powered millisecond pulsars indicates that their initial spins are set by spin-up to accretion spin equilibrium. Many are well below the spin-equilibrium line for accretion rates $\sim 10^{-2} {\dot M}_E$ but have very long magnetic-braking spin-down times, indicating that they had accretion rates $\ll {\dot M}_E$ when accretion ended or never reached spin equilibrium. 

After more than two decades of effort, accreting MSPs in LMXBs have at last been found. The discovery of these stars promises important advances in our understanding of the evolution of neutron stars in LMXBs and the formation of radio-emitting, rotation-powered MSPs.

We thank L. Bildsten, D. Chakrabarty, P. Kaaret, M. van der Klis, M.~C. Miller, D. Psaltis, and J. Swank for helpful discussions. This research was supported in part by NASA grants NAG~5-12030 and NAG~5-8740, NSF grant AST~0098399, and the funds of the Fortner Endowed Chair at the University of Illinois.

\begin {references}

\reference
Alpar, M.~A., Cheng, A.~F., Ruderman, M.~A., \& Shaham, J. 1982, Nature, 300, 178

\reference
Arzoumanian, Z., Cordes, J.M., \& Wasserman, I. 1999, ApJ 520, 696

\reference
Backer, D.~C., Kulkarni, S.~R., Heiles, C., Davis, M.~M., \& Goss, W.~M. 1982,  Nature, 300, 615 

\reference
Bhattacharya, D.  1995, in X-Ray Binaries, ed.\ W.~H.~G. Lewin, J. van Paradijs \& E.~P.~J van den Heuvel (Cambridge, Cambridge University Press), 233 

\reference
Bhattacharya, D., \& Srinivasan, G.  1995, in X-Ray Binaries, ed.\ W.~H.~G. Lewin, J. van Paradijs \& E.~P.~J van den Heuvel (Cambridge, Cambridge University Press), 495 

\reference
Bhattacharya, D., \& van den Heuvel, E.~P.~J. 1991, Phys.\ Rep., 203, 1

\reference
Bildsten, L. 1998, \apj, 501, L89

\reference
Chakrabarty, D., \& Morgan, E. H. 1998, Nature, 394, 346

\reference
Chakrabarty, D., Morgan, E.~H., Muno, M.~P., Galloway, D.~K., Wijnands, R., van der Klis, M., \& Markwardt, C.~B. 2003, Nature, 424, 42

\reference
Elsner, R.~F., Ghosh, P., \& Lamb, F.~K. 1980, \apj, 241, L55

\reference
Galloway, D.~K., Chakrabarty, D., Morgan, E. H., \& Remillard, R. A. 2002, ApJ, 576, L137

\reference
Galloway, D.~K., Chakrabarty, D.,  Muno, M.~P., \& Savov, P. 2001, ApJ, 549, L85

\reference
Ghosh, P., \& Lamb, F.~K.  1979, \apj, 234, 296 (GL79)

\reference
Ghosh, P., \& Lamb, F.~K. 1992, in X-Ray Binaries and Recycled Pulsars, ed.\ E.~P.~J. van den Heuvel \& S. Rappaport (Dordrecht: Kluwer), 487 (GL92)

\reference
Hartman, J. M., Chakrabarty, D., Galloway, D. K.,  Muno, M. P., Savov, P., Mendez, M., van Straaten, S., Di Salvo, T., 2003, American Astronomical Society, HEAD Meeting No.~35, abstract~17.38

\reference
Hasinger, G., \& van der Klis, M.  1989, \aap, 186, 153

\reference
Hobbs, G., Lyne, A.~G., Kramer, M., Martin, C.~E., \& Jordan, C.~A. 2004, \mnras, submitted

\reference
Hobbs, G.~B., \& Manchester, R.~N. 2004, ATNF Pulsar Catalogue, v1.2, \break http://www.atnf.csiro.au/research/pulsar/psrcat/psrcat\_help.html

\reference
Kaaret, P., Zand, J. J. M. in't., Heise, J., \& Tomsick, J. A., 2002, ApJ, 575, 1018

\reference
Kaaret, P., Zand, J. J. M. in't., Heise, J., \& Tomsick, J. A. 
2003, ApJ, 598, 481

\reference
Lamb, F.~K. 1989, in Proc.\ 23rd ESLAB Symp.\ on X-ray Astronomy, ed. J. Hunt \& B. Battrick (ESA SP-296), 215

\reference
Lamb, F.~K. 2003, in X-Ray Binaries and Gamma-Ray Bursts, ed.\ E.~P.~J. van den Heuvel, L. Kaper, E. Rol \&  Ralph A.~M.~J. Wijers (San Francisco: Astron. Soc. Pacific, 2003), 221

\reference
Lamb, F.~K. 2005, in Electromagnetic Emission of Neutron Stars, ed.\ A. Baykal et al.\ (Dordrecht: Kluwer), in press

\reference
Lamb, F.~K., \& Miller, M.~C. 2001, \apj, 554, 1210

\reference
Lamb, F.~K., \& Miller, M.~C. 2004, \apj, submitted, astro-ph/0308179

\reference
Lamb, F.~K., \& Yu, W. 2004, in preparation

\reference
Lewin, W.~H.~G., van Paradijs, J., \& Taam, R. 1995, in X-Ray Binaries, ed.\ W.~H.~G. Lewin, E.~P.~J. van den Heuvel \& J. van~Paradijs (Cambridge: Cambridge Univ. Press), 175

\reference
Lindblom, L., \& Owen, B. 2002, Phys.\ Rev. D, 65, 063006

\reference
Markwardt, C.~B., Smith, E., \& Swank, J.~H. 2003a, IAU Circ.\ 8080

\reference
Markwardt, C.~B., Strohmayer, T.~E.  \& Swank, J.~H. 1999, ApJ, 512, L125 

\reference
Markwardt, C.~B., \& Swank, J.~H. 2003b, IAU Circ., 8144, 1

\reference
Markwardt, C.~B., Swank, J.~H., Strohmayer, T.~E., in 't Zand, 
J.~J.~M., \& Marshall, F.~E. 2002, ApJ, 575, L21

\reference
Markwardt, C.~B., et al. 2004, in preparation

\reference
Miller, M.~C., Lamb, F.~K., \& Psaltis, D. 1998, \apj, 508, 791 (MLP98)

\reference
Muno, M~P., Chakrabarty, D., \& Galloway, D.~K., \& Psaltis, D. 2002, \apj, 580, 1048

\reference
Muno, M.~P., Fox, D.~W., Morgan, E.~P. 2000, \apj, 542, 1016

\reference
Psaltis, D., \& Chakrabarty, D. 1999, \apj, 521, 332

\reference
Psaltis, D., \& Lamb, F. K. 1998, in Neutron Stars and Pulsars, ed.\ N. Shibazaki, N. Kawai, S. Shibata, \& T. Kifune (Tokyo: Univ. Acad. Press), 179

\reference
Radhakrishnan, V., \& Srinivasan, G. 1982, Curr.\ Sci., 51, 1096

\reference
Ritter H., \& King A.~R. 2001, in ASP Conf.\ Ser.\ Vol.~229, Evolution of Binary and Multiple Star Systems, ed.\ Podsiadlowski, P., Rappaport S.~A., King A.~R., D'Antona F. \& Burderi L. (San Francisco, Astron.\ Soc.\ Pac.), 423

\reference
Shibazaki, N., et al.\ 1989, Nature, 342, 656

\reference
Smith, D.~A., Morgan, E.~H., \& Bradt, H. 1997,  ApJ, 482, L65

\reference
Strohmayer, T.~E., Jahoda, K., Giles, A.~B., \& Lee, U.  1997, ApJ, 486, 355

\reference
Strohmayer, T.~E., \& Markwardt, C.~M. 2002, \apj, 577, 337

\reference
Strohmayer, T.~E., Markwardt, C.~M., Swank, J.~H., \& in't Zand,
J.J.M. 2003, ApJ, 596, L67

\reference
Strohmayer, T.~E., Zhang, W., Swank, J.~H., Smale, A.~P., 
Titarchuk, L., \& Day, C. 1996, ApJ, 469, L9

\reference 
Strohmayer, T.~E., Zhang, W., Swank, J.~H., White, N.~E., \& Lapidus, I. 1998, ApJ, 498, L135

\reference
Ushomirsky, G., Cutler, C., \& Bildsten, L. 2000, \mnras,         319, 902

\reference
van den Heuvel, E.~P.~J. 1992, in X-Ray Binaries and Recycled Pulsars, ed.\ E.~P.~J. van den Heuvel \& S.~A. Rappaport (Dordrecht: Kluwer), 233.

\reference
White, N., \& Stella, L. 1987, \mnras, 231, 325

\reference
Wijnands, R., Strohmayer, T., Franco, L.~M., 2001,  ApJ, 554, L59

\reference
Wijnands, R., \& van der Klis, M. 1998, Nature,  394, 344

\reference
Wijnands, R., van der Klis, M., Homan, J., Chakrabarty, D.,
Markwardt, C.B., \& Morgan, E.~H. 2003, Nature, 424, 44

\reference
Zhang, W., Jahoda, K., Kelley, R.~L., Strohmayer, T.~E., Swank, J.~H. \& Zhang, S.~N. 1998, ApJ, 495, L9  

\reference
Zhang, W., Lapidus, I., Swank, J.~H., White, N.~E., \&  Titarchuk, L. 1996, IAU Circ. 6541

\end {references}

\end {document}